\documentclass{book}
\usepackage{Wiley-AuthoringTemplate}

\begin{document}

\title{Quantum simulation of clustered photosynthetic light harvesting in a superconducting quantum circuit}

\author[1,2]{Ming-Jie Tao}
\author[3]{Ming Hua}
\author[1]{Na-Na Zhang}
\author[1]{Wan-Ting He}
\author[1]{Qing Ai}
\author[1,4]{Fu-Guo Deng}

\address[1]{Department of Physics, Applied Optics Beijing Area Major Laboratory, Beijing Normal University, Beijing, 100875, China}
\address[2]{Faculty of foundation, Space Engineering University, Beijing, 101416, China}
\address[3]{Department of Applied Physics, School of Physical Science and Technology, Tianjin Polytechnic University, Tianjin, 300387, China}
\address[4]{NAAM-Research Group, Department of Mathematics, Faculty of Science, King Abdulaziz University, Jeddah, 21589, Saudi Arabia}
\address*{Corresponding author: Qing Ai; aiqing@bnu.edu.cn}

\maketitle%

\begin{abstract}{}
We propose a method to emulate the exciton energy transfer (EET) of photosynthetic complexes
in a quantum superconducting circuit. Our system is composed of two pairs of superconducting charge qubits coupled to two separated high-$Q$ superconducting transmission line resonators (TLRs), respectively. The two TLRs interact with each other capacitively. When the frequencies of the qubits are largely detuned from those of the TLRs, we simulate the process of EET from the first qubit to the fourth qubit. By tuning the couplings between the qubits and the TLRs, as well as the coupling between the two TLRs, we can modify the effective coupling strengths between the qubits and thus study the geometric effects on the EET. It is shown that a moderately-clustered geometry supports optimal EET by using exciton delocalization and an energy matching condition. And the population loss
during the EET has been trapped in the two TLRs.
\end{abstract}


\section{Introduction}
\label{sec:Introduction}

Energy plays an important role in modern society. The chemical energy supporting all lives on earth is mainly from the solar energy harvested by photosynthesis \cite{Cheng09,Lambert13,Chin13,Blankenship02}.
The solar energy can be captured and transferred to the reaction centers of photosynthetic systems in a short time with high efficiency \cite{Cheng09,Cheng06,Engel07}. Therefore, it might be beneficial to learn from the natural photosynthesis to design efficient artificial light-harvesting devices.

In the past few decades, many researches were focused on the study of the exciton energy transfer
(EET) process in photosynthesis \cite{Engel07,Amerongen00,Knox96,Leegwater96,Savikhin97,Mancal13}.
Based on the quantum dynamics of open systems \cite{Lambert13,Cao20}, much knowledge has been learned about the efficiency of the EET \cite{Jang14,Fassioli12,Collini10,Ai14,Hildner13}, together with the spatial and energetic arrangement of the pigments \cite{Sener11,Scholak11-2,Marin11,Cleary13,Xu18}.
The experimentally-observed coherent phenomena in 2D spectroscopy may be attributed to nearly-resonant coupling to an underdampped vibrational mode in the bath \cite{Cao20,Mourokh15}.
In natural photosynthesis, EET can be accomplished within $100$ picoseconds with almost $100\%$
efficiency \cite{Fleming94}. Schulten \emph{et al.} \cite{Hu97} observed that the bacteriochlorophylls involved
in the overall excitation transfer are found in a coplanar arrangement. Ishizaki and
Fleming \cite{Ishizaki09} showed that, by dimerization, the energy flow in the Fenna-Mattthew-Olson (FMO) complex occurs
primarily through two EET pathways. Yang \emph{et al.} \cite{Yang10} found out that the dimerization
in light-harvesting complex II (LH2) can effectively speed up the energy transfer between LH2 rings
due to symmetry breaking. In a wheel-shaped artificial light-harvesting complex, the energy of the initially-excited antenna can be efficiently directed to the reaction center, with quantum beating lasted for hundreds of femtoseconds \cite{Ghosh11-1,Ghosh11-2}.
In 2013, Ai \emph{et al.} \cite{Ai13} revealed that clustered geometry
utilizes exciton delocalization and energy matching to optimize EET, which can be utilized to explain the efficient EET in FMO complex \cite{Chen13}.

In order to obtain the explicit relation between geometry and efficiency, some researches have explored
random networks to discover some interesting findings \cite{Mostarda13,Knee17,Zech14}. Because the EET
is sensitively influenced by the interference within the photosynthetic complex, even small changes in
the geometry of the complex could turn constructive interference to destructive and thus result in
significant drop in the efficiency. Pair sites renders EET properties robust against perturbations \cite{Mostarda13}.
Compact structures tend to display high performance in the transport dynamics \cite{Knee17}. The networks
characterized by Hamiltonians with centrosymmetry outperform those with completely-random arrangements \cite{Zech14}.
These discoveries seem to suggest that clustered geometries could favour efficient quantum transport.

On the other hand, much progress has been made in quantum information science,
inspiring several interesting quantum simulation \cite{Buluta09,Georgescu14} experiments to verify the design principles for optimal light-harvesting \cite{Ghosh11-1,Ghosh11-2,Wang18,Zhang20,Mostame12,Potocnik17,Chin18,Rey13}. By using bath engineering and the gradient ascent pulse engineering algorithm, Wang \emph{et al.} \cite{Wang18} performed an experimental quantum simulation of photosynthetic energy transfer by using nuclear magnetic resonance (NMR). It was demonstrated that the open quantum dynamics in an $N$-level system, with arbitrary Hamiltonian and bath spectral density, can be effectively emulated by an NMR system with $\log_{2}N$ qubits \cite{Zhang20}. Meanwhile,
Gorman \emph{et al.} \cite{Gorman18} showed in a trapped-ion system that the long-lived vibrational mode in the bath
can assist the energy transfer. Superconducting quantum circuits provide another intriguing
platform for quantum simulation \cite{Blais04,You05,Blais07,Leek07,Ai09,You11,Xiang13,Gu17}. In 2012, Mostame \emph{et al.} \cite{Mostame12}
simulated a complicated environment with a given spectral density for the EET in photosynthetic complexes
by using inductor-resistor-capacitor oscillators. In 2018, Poto\v{c}nik \emph{et al.} \cite{Potocnik17,Chin18}
experimentally demonstrated that light harvesting for a given geometry can be optimized by tuning
the environmental noise.

However, although they have shown the potential of optimizing energy transfer by engineering the bath,
none of them have experimentally demonstrated the effect of geometry on the EET efficiency. In
Ref.~\cite{Ai13}, it was shown that, in a linear geometry, moderate dimerization promotes the energy
transfer. Therefore, it might be interesting to simulate the EET in different geometries to verify
the design principals of optimal geometries. In this paper, we design a system composed of four
superconducting charge qubits and two superconducting transmission line resonators (TLRs). Here,
two qubits form a pair and are coupled to one TLR. And the two TLRs are capacitively coupled with
each other. Although there are no direct interactions between the qubits, the effective couplings
among them can be induced by the simultaneous couplings to the common mode in each TLR  \cite{Yang13,Yang16}. Furthermore,
the effective couplings can be tuned by adjusting the level spacings of the qubits, and their
interaction strengths with the TLRs \cite{Allman10}, and the coupling strength between the two TLRs. In this way,
we can investigate the EET for different geometries. Alternatively, for the sake of simplicity,
the direct couplings between the qubits can be introduced instead. This alternative setup can
simplify the arrangement and approximately simulate the EET dynamics for the parameter regime when
the nearest-neighbour couplings dominate the transport.

This paper is organized as follows: In the next section, we briefly introduce the theory for
describing the EET in photosynthesis. In Sec.~\ref{sec:scheme}, we propose a setup consisting
of four superconducting charge qubits and two TLRs to simulate the photosynthetic energy transfer.
The effective Hamiltonian for the four qubits is obtained by the Fr\"{o}hlich-Nakajima transformation.
In Sec.~\ref{sec:simulation}, the energy transfer dynamics is numerically simulated by solving
the Lindblad master equation, which confirms the previous investigation in Ref.~\cite{Ai13}. Finally,
the experimental feasibility and the main conclusions are discussed in Sec.~\ref{sec:summary}.

\section{Photosynthetic Light Harvesting}
\label{sec:EET}

In the photosynthesis with 4 chromophores, the EET is governed by the Frenkel-exciton Hamiltonian \cite{Cheng09,Ai13}
\begin{equation}
H=\sum_{j=1}^{4}\varepsilon_j\vert j\rangle\langle j\vert+\sum_{i\neq j=1}^{4}J_{ij}\vert i\rangle\langle j\vert+\textrm{h.c.},
\end{equation}
where $\varepsilon_j$ is the site energy when $j$-th chromophore is in the excited state,
$\vert j\rangle$ is the state when $j$-th chromophore is in the excited state while all other
chromophores are in the ground state, $J_{ij}$ is the dipole-dipole interaction between $i$-th
and $j$-th chromophores.

Due to the strong coupling $J_{ij}\gtrsim\vert\varepsilon_i-\varepsilon_j\vert$,
the exciton energy can coherently oscillate between any two sites $i$ and $j$. However,
because of the pure-dephasing-form system-bath Hamiltonian,
\begin{equation}
H_{SB}=\sum_{j,k}g_{jk}\vert j\rangle\langle j\vert(a^\dagger_{jk}+a_{jk}),
\end{equation}
where $g_{jk}$ is the coupling strength between $j$-th chromophore and its local bath mode
with frequency $\omega_k$ and creation (annihilation) operator $a^\dagger_{jk}$ ($a_{jk}$),
the exciton energy can be irreversibly transferred to the target chromophore.
In general, the system-bath coupling is described by the spectral density,
\begin{equation}
G(\omega)=\sum_{k}g_{jk}^2\delta(\omega-\omega_k).
\end{equation}

For a given geometry, the position $\vec{r}_j$ and transition dipole $\vec{\mu}_j$
of every chromophore is fixed and thus the dipole-dipole interaction between
any pair of two chromophores is determined by
\begin{equation}
J_{ij}=\frac{1}{4\pi\varepsilon_0r_{ij}^3}[\vec{\mu}_i\cdot\vec{\mu}_j-3(\vec{\mu}_i\cdot\hat{r}_{ij})
(\vec{\mu}_j\cdot\hat{r}_{ij})],
\end{equation}
where $\vec{r}_{ij}=r_{ij}\hat{r}_{ij}=\vec{r}_{i}-\vec{r}_{j}$
is the displacement vector from site $j$ to site $i$, $\varepsilon_0$ is the vacuum
permittivity. In addition, the spatial distribution of $\varepsilon_j$ also facilitates
the energy transfer by making use of the energy gradient towards the target chromophore,
as shown in Fig.~\ref{figure1}(b). In Ref.~\cite{Ai13}, by using the coherent modified
Redfield theory \cite{Ai14,Hwang15,Chang15}, it is shown that, in a clustered geometry,
exciton delocalization and energy matching cooperate to optimize EET.

\section{Physical setup}
\label{sec:scheme}

\begin{figure}[!tpb]   %  figure1
\begin{center}
\includegraphics[width=8 cm]{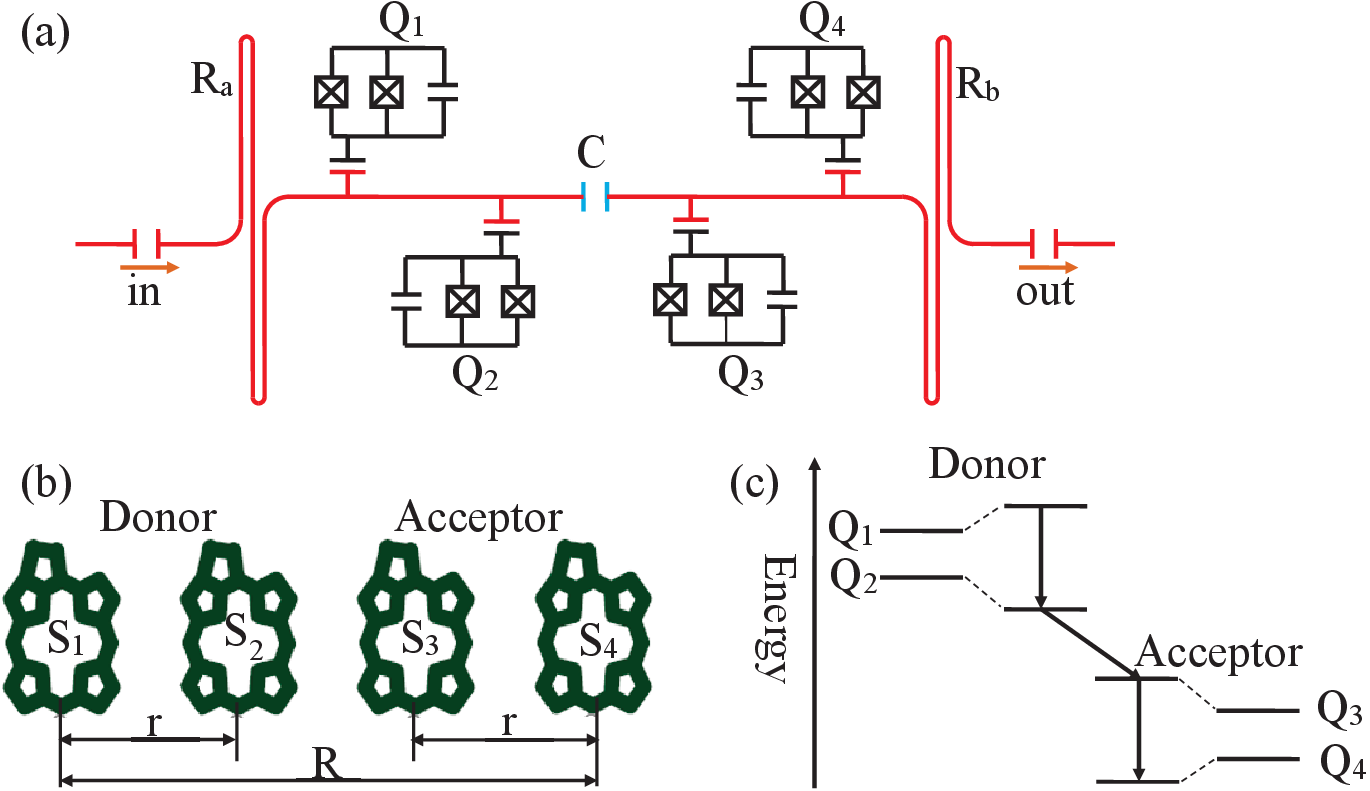}
\caption{(a) Schematic diagram of the superconducting circuit for simulating
photosynthetic energy transfer. The two TLRs are capacitively coupled with each other.
Charge qubits $Q_1$ and $Q_2$ form the donor, while qubits $Q_3$ and $Q_4$ act as the
acceptor. $Q_1$ and $Q_2$ ($Q_3$ and $Q_4$) are capacitively coupled to TLR $R_{a}$
($R_b$). (b) In a linear photosynthetic system, sites 1 and 2 form the donor pair, while sites 3 and 4 are the acceptor pair. The total distance between the two ends is fixed at $R$ and the intra-pair distance is $r$.
(c) Energy configuration of the four qubits for energy transfer, where only the excited states are involved in the EET and the ground states are not shown for simplicity.
}\label{figure1}
\end{center}
\end{figure}

Let us consider a superconducting quantum circuit composed of four superconducting charge
qubits and two $1$D high-$Q$ superconducting TLRs, as shown in Fig.~\ref{figure1}(a).
 In order to study the cluster-to-cluster geometric effects, there should be two
clusters at least, and two qubits can form a cluster. Therefore, four qubits are
the minimum number of qubits to observe this effect. As shown in Fig.~\ref{figure1}(b),
we investigate the EET efficiency in a linear photosynthetic complex with 4 chromophores.
The distance between the two ends is fixed at $R=40~{\AA}$, while sites 1 and 2
form the donor pair, and sites 3 and 4 are the acceptor pair, with intra-pair distance $r<R/3$.
The energy-level diagram of the four qubits is schematically shown in Fig.~\ref{figure1}(c).
Qubits $Q_{1}$ and $Q_{2}$ ($Q_{3}$ and $Q_{4}$) are coupled to the TLR $R_{a}$ ($R_{b}$)
capacitively. Here, we take $Q_{1}$ and $Q_{2}$ as donors because their energies are higher
than those of the qubits $Q_{3}$ and $Q_{4}$, acting as acceptors. The effective couplings
among these qubits exhibit the geometrical effects in photosynthetic complexes, because
the couplings between pigments sensitively depend on their relative distances and orientations
of electric dipoles. The couplings other than the nearest-neighbor couplings subtly modify
the energy spectrum and thus the quantum dynamics. Furthermore, instead of direct couplings
among qubits, we simplify the quantum circuit by introducing the additional TLRs. In other
words, the TLRs play the role as the quantum data bus to induce the indirect couplings
between the qubits. The frequencies of TLRs should be much smaller than the qubits to
avoid the excitation of the TLRs. The distances between any two qubits are far enough
to avoid direct interactions between them. Therefore, the energy is transferred from $Q_{1}$ to $Q_{4}$ by
the indirect interactions among the qubits induced by simultaneously couplings to the
common TLRs.

Under the rotating-wave approximation \cite{Ai10}, the Hamiltonian of the four qubits
and two TLRs can be written as
\begin{eqnarray}         %  equation 1
H_{1}&=&\omega_{a}a^{\dagger}a+\sum_{j_1=1}^{2}\left[\frac{\omega_{j_1}}{2}\sigma_{j_1}^{z}+g_{j_1}\left(a^{\dagger}\sigma_{j_1}^{-}+a\sigma_{j_1}^{+}\right)\right]            \nonumber\\
&& + \omega_{b}b^{\dagger}b+\sum_{j_2=3}^{4}\left[\frac{\omega_{j_2}}{2}\sigma_{j_2}^{z}+g_{j_2}\left(b^{\dagger}\sigma_{j_2}^{-}+b\sigma_{j_2}^{+}\right)\right]          \nonumber\\
&& + g_{a}^{b}\left(a^{\dagger}+a\right)\left(b^{\dagger}+b\right),
\label{eq1}
\end{eqnarray}
where $\omega_{a}$, $\omega_{b}$, $\omega_{j_{1}}$, and $\omega_{j_{2}}$ are the
transition frequencies of the TLRs $R_{a}$ and $R_{b}$, and qubits $Q_{j_{s}}$ ($s=1,2$),
respectively. Here, $j_1=1,2$ and $j_2=3,4$. $g_{j_{1}}$ ($g_{j_{2}}$) is the coupling
strength between the qubit $Q_{j_{s}}$ ($Q_{j_{2}}$) and the TLR $R_{a}$ ($R_{b}$).
$g_{a}^{b}$ is the coupling strength between TLRs $R_a$ and $R_b$. $a^{\dagger}$
and $b^{\dagger}$ are the creation operators of $R_{a}$ and $R_{b}$, respectively.
$\sigma_{j_{s}}^{+}=\vert e\rangle_{j_{s}}\langle g\vert$ and $\sigma_{j_{s}}^{z}$
are the rising and Pauli operator of $Q_{j_{s}}$, respectively. $|g\rangle_{j_{s}}$
and $|e\rangle_{j_{s}}$ are the ground and excited states of $Q_{j_{s}}$, respectively.

By using the Fr\"{o}hlich-Nakajima transformation \cite{Blais07}
\begin{eqnarray}\label{eq2}    %  equation 2
U&=&\exp\left[\sum_{j_1=1}^{2}\frac{g_{j_{1}}}{\delta_{j_{1}}}\left(a^{\dagger}\sigma_{j_{1}}^{-}-a\sigma_{j_1}^{+}\right)
+\sum_{j_2=3}^{4}\frac{g_{j_2}}{\delta_{j_2}}\left(b^{\dagger}\sigma_{j_2}^{-}-b\sigma_{j_2}^{+}\right)\right]
\end{eqnarray}
with $\delta_{j_{1}}=\omega_{j_{1}}-\omega_{a}\gg g_{j_{1}}$ and
$\delta_{j_{2}}=\omega_{j_{2}}-\omega_{b}\gg g_{j_{2}}$, the original
Hamiltonian of the system $H_{1}$ becomes
\begin{eqnarray}\label{eq3}     %  equation 3
H_{2}&=&U^{\dagger}H_{1}U.
\end{eqnarray}
In the appendix, we give the detailed expression of $H_2$.

When the TLRs are initially prepared in the vacuum state and the high-order terms of $g_{j_{s}}/\delta_{j_{s}}$ can be omitted, the Hamiltonian $H_2$ can be reduced to
\begin{eqnarray}\label{eq5}         %  equation 5
H_{\rm{eff}}&=&\sum_{j_1=1}^{2}\left(\omega_{j_1}+\frac{g_{j_1}^{2}}{\delta_{j_1}}\right)\vert e\rangle_{j_1}\langle e\vert
        +\sum_{j_2=3}^{4}\left(\omega_{j_2}+\frac{g_{j_2}^{2}}{\delta_{j_2}}\right)\vert e\rangle_{j_2}\langle e\vert       \nonumber\\
&&  +J_{12}\left(\sigma_{1}^{-}\sigma_{2}^{+}+\sigma_{1}^{+}\sigma_{2}^{-}\right)+J_{34}\left(\sigma_{3}^{-}\sigma_{4}^{+}+\sigma_{3}^{+}\sigma_{4}^{-}\right)   \nonumber\\
&&  +J_{23}\left(\sigma_{2}^{-}\sigma_{3}^{+}+\sigma_{2}^{+}\sigma_{3}^{-}\right)+J_{13}\left(\sigma_{1}^{-}\sigma_{3}^{+}+\sigma_{1}^{+}\sigma_{3}^{-}\right) \nonumber\\
&&  +J_{24}\left(\sigma_{2}^{-}\sigma_{4}^{+}+\sigma_{2}^{+}\sigma_{4}^{-}\right)+J_{14}\left(\sigma_{1}^{-}\sigma_{4}^{+}+\sigma_{1}^{+}\sigma_{4}^{-}\right),
\end{eqnarray}
where $J_{12}=\frac{g_{1}g_{2}}{2\delta_{1}\delta_{2}}\left(\delta_{1}+\delta_{2}\right)$,
$J_{34}=\frac{g_{3}g_{4}}{2\delta_{3}\delta_{4}}\left(\delta_{3}+\delta_{4}\right)$,
$J_{23}=\frac{g_{a}^{b}g_{2}g_{3}}{\delta_{2}\delta_{3}}$, $J_{13}=\frac{g_{a}^{b}g_{1}g_{3}}{\delta_{1}\delta_{3}}$,
$J_{24}=\frac{g_{a}^{b}g_{2}g_{4}}{\delta_{2}\delta_{4}}$, and $J_{14}=\frac{g_{a}^{b}g_{1}g_{4}}{\delta_{1}\delta_{4}}$
are the indirect coupling strengths between any two qubits, respectively.
In order to mimic the geometric effects, the effective couplings $J_{ij}$'s should be tunable. This can be achieved by adjusting detunings $\delta_j$'s, qubit-resonator couplings, and resonator-resonator coupling \cite{Allman10}.

When a quantum system interacts with a quantum bath, the states of the quantum system
entangle with those of the quantum bath \cite{Mancal13}. The correlation function of
the classical bath is real-valued and time-symmetric. However, when a quantum system interacts
with a quantum bath, the quantum system experiences time-dependent transition frequencies
and thus results in a complex-valued time-asymmetric correlation function of the bath \cite{Mancal13}.
Due to this difference, the energy generally prefers to transfer from the higher-energy level
to the lower-energy level of the quantum system interacting with a quantum bath \cite{Mancal13}.
Furthermore, the populations of all levels tend to be equal at the steady state in the case
of a classical bath. The latter has been recently observed in the experimental NMR simulation \cite{Wang18}.
In our configuration, the transition frequencies of the four qubits are assumed to satisfy
the following relation, i.e., $\left(\omega_{1}+\frac{g_{1}^{2}}{\delta_{1}}\right)>\cdots>\left(\omega_{4}+\frac{g_{4}^{2}}{\delta_{4}}\right)$,
as shown in Fig.~\ref{figure1}(b). Moreover, coupling strengths $J_{12}$ and $J_{34}$ are
assumed to be larger than $J_{23}$ to indicate that qubits $Q_1$ and $Q_2$ form the donor
pair, while qubits $Q_3$ and $Q_4$ are the acceptor pair. For simplicity, we take
$g_{1}=g_{4}$, $g_{2}=g_{3}$ and $\delta_{1} \approx \delta_{2} \approx \delta_{3} \approx \delta_{4}$
to achieve $J_{12}=J_{34}>J_{23}$ \cite{Ai13}. Thus we use the following parameters
$\omega_{a}/2\pi=\omega_{b}/2\pi=3$~GHz, $\omega_{1}/2\pi=13.115$~GHz,
$\omega_{2}/2\pi=13.009$~GHz, $\omega_{3}/2\pi=12.991$~GHz, $\omega_{4}/2\pi=13.078$~GHz \cite{Blais04,Blais07}
in our numerical simulations. The proposed parameters in the superconducting circuit model
can be obtained from those found in biological systems by scaling down a factor $3\times10^4$.

\section{simulation of EET process}
\label{sec:simulation}

In the previous section, we obtain the effective Hamiltonian for the 4 qubits by the
Fr\"{o}hlich-Nakajima transformation and the rotating-wave approximation. It is reasonable
to question the validity of rotating-wave approximation since there is strong coupling
between the TLRs and the qubit-TLR couplings are relative large as compared to the
frequencies of the TLRs. Hereafter, we shall numerically simulate quantum dynamics
of the master equation under the exact Hamiltonian without the rotating-wave approximation.

In the interaction picture with respect to
\begin{eqnarray}            %  equation 6
H_{0}&=&\omega_{a}a^{\dagger}a+\sum_{j_1=1}^{2}\frac{\omega_{j_1}}{2}\sigma_{j_1}^{z}+\omega_{b}b^{\dagger}b+\sum_{j_2=3}^{4}\frac{\omega_{j_2}}{2}\sigma_{j_2}^{z},
\end{eqnarray}
we can derive the Lindblad-form master equation as \cite{Breuer02,Ai14}
\begin{eqnarray}        %  equation 9
\dot{\rho}&=&-i\left[H_{I},\rho\right]+\sum_{r=a,b}\kappa_{r}\left(N_{r}+1\right)D\left[r\right]\rho         \nonumber\\
&&+\sum_{r=a,b}\kappa_{r}N_{r}D\left[r^{\dagger}\right]\rho+\sum_{l=1}^{4}\Gamma_{l}^{\gamma}\left(N_{l}+1\right)D\left[\sigma_{l}^{-}\right]\rho     \nonumber\\
&&+\sum_{l=1}^{4}\Gamma_{l}^{\gamma}N_{l}D\left[\sigma_{l}^{+}\right]\rho+\sum_{l=1}^{4}\Gamma_{l}^{\phi}D\left[\sigma_{l}^{z}\right]\rho,
\end{eqnarray}
where the interaction Hamiltonian reads
\begin{eqnarray}            %  equation 8
H_\textrm{I}&=&\sum_{j=1}^{2}g_{j}\left(a\sigma_{j}^{+}e^{i\delta_{j}t}+a^{\dagger}\sigma_{j}^{+}e^{i(\omega_{a}+\omega_{j})t}\right)           \nonumber\\
&&+\sum_{j=3}^{4}g_{j}\left(b\sigma_{j}^{+}e^{i\delta_{j}t}+b^{\dagger}\sigma_{j}^{+}e^{i(\omega_{b}+\omega_{j})t}\right)\nonumber\\
&&+g_{a}^{b}\left(abe^{-i\Delta_{a}^{b}t}+ab^{\dagger}e^{i\delta_{a}^{b}t}\right)+\textrm{h.c.} ,
\end{eqnarray}
\begin{eqnarray}           %  equation 10,11,12
\delta_{a}^{b}&=&\omega_{b}-\omega_{a},\\
\Delta_{a}^{b}&=&\omega_{b}+\omega_{a},\\
D[A]\rho&=&\left(2A\rho A^{\dagger}-A^{\dagger}A\rho-\rho A^{\dagger}A\right)/2,      \\
N_{r}&=&\frac{1}{\exp\left(\hbar\omega_{r}/k_{B}T\right)-1},         \\
N_{l}&=&\frac{1}{\exp\left(\hbar\omega_{l}/k_{B}T\right)},
\end{eqnarray}
$\kappa_{r}~(r=a,b)$ is the leakage rate of TLR $r$,
$\Gamma_{l}^{\gamma}$ and $\Gamma_{l}^{\phi}$ ($l=1,2,3,4$) are the spontaneous emission
and pure-dephasing rates of the $l$-th qubit, respectively.

\begin{figure}[!h]%[tpb]  % figure 2
\begin{center}
\includegraphics[width=8.5 cm,angle=0]{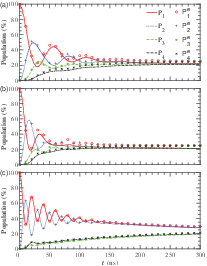}
\caption{ The curves (symbol) show the propagation of the populations of excitation on each of the four qubits by $H_\textrm{I}$ ($H_\textrm{eff}$) for three different geometries: (a) equally-spaced geometry with $J_{12}=J_{34}\approx J_{23}$ and $r\approx R/3$, (b) moderately-clustered geometry with $J_{12}=J_{34}=1.62J_{23}$ and $r<R/3$, (c) over-clustered geometry with $J_{12}=J_{34}=3.11J_{23}$ and $r\ll R/3$. The red solid line ($\circ$) is for $P_{1}$ ($P_{1}^\textrm{e}$), blue dotted line ($+$) for $P_{2}$ ($P_{2}^\textrm{e}$), green dashed line ($*$) for $P_{3}$ ($P_{3}^\textrm{e}$), and black dash-dotted line ($\times$) for $P_{4}$ ($P_{4}^\textrm{e}$).
}\label{figure2}
\end{center}
\end{figure}

In natural photosynthesis, the energy transfer is generally restricted in the single-excitation
subspace. For simplicity, we label the bases as
$\vert1\rangle=|e\rangle_1|g\rangle_2|g\rangle_3|g\rangle_4\vert0_a0_b\rangle$,
$\vert2\rangle=|g\rangle_1|e\rangle_2|g\rangle_3g\rangle_4\vert0_a0_b\rangle$,
$\vert3\rangle=|g\rangle_1|g\rangle_2|e\rangle_3|g\rangle_4\vert0_a0_b\rangle$,
and $\vert4\rangle=|g\rangle_1|g\rangle_2|g\rangle_3|e\rangle_4\vert0_a0_b\rangle$.
Here, $\vert n_an_b\rangle$ is the Fock state of $\mathrm{TL}\mathrm{R}_{a}$ and $\mathrm{TL}\mathrm{R}_{b}$.
In addition, $\vert a\rangle=|g\rangle_1|g\rangle_2|g\rangle_3|g\rangle_4\vert 1_a0_b\rangle$
and $\vert b\rangle=|g\rangle_1|g\rangle_2|g\rangle_3|g\rangle_4\vert 0_a1_b\rangle$
indicate single-excitation in one of the TLRs. In our simulation, the system composed of
four superconducting qubits and two TLRs is initially prepared at the state $\vert1\rangle$.
In Fig.~\ref{figure2}, we show the time evolution of the populations of single-excitation
states of the four qubits $P_{m}=\langle m\vert\rho\vert m\rangle$ ($m=1,2,3,4$). In a
realistic experiment, each qubit can be dispersively coupled to a TLR.  The qubit's
population can be extracted by measuring the phase of the signal  in the TLR \cite{Kim11}.
And the qubit's decay time will not be significantly  modified by Purcell-like couplings
to the TLRs, which will be discussed in  detail in Sec.~\ref{sec:summary}. Here, we take
$1/\Gamma_{j}^{\gamma}=3$~$\mu$s and $1/\Gamma_{j}^{\phi}=70$~ns ($j=1,2,3,4$) \cite{Kim11},
which meet the requirement $\Gamma_{j}^{\phi}\gg\Gamma_{j}^{\gamma}$ \cite{Plenio08,Rebentrost09,Chin10}.
The leakage rates of two TLRs are $\kappa_a^{-1}=\kappa_b^{-1}=10$~$\mu$s \cite{Wallraff05}.

In Fig.~\ref{figure2}, we demonstrate the population dynamics of the four qubits for three
different geometries, corresponding to three different sets of nearest-neighbor couplings
$J_{12}$, $J_{34}$, and $J_{23}$. In Ref.~\cite{Ai13}, it has been proven that the next-nearest-neighbor
couplings $J_{13}$ and $J_{24}$, and the end-to-end coupling $J_{14}$ plays a minor role
in the EET. First of all, we would like to simulate the energy transfer for an equal-coupling
geometry. Therefore, we investigate the energy transfer dynamics of the system with approximately
equal-coupling strengths between adjacent qubits in Fig.~\ref{figure2}(a). It corresponds
to the equally-spaced geometry in Ref.~\cite{Ai13}. In order to achieve $J_{12}=J_{34}\approx J_{23}$,
coupling strengths are assumed to be $g_{1}/2\pi=g_{4}/2\pi=100$~MHz, $g_{2}/2\pi=g_{3}/2\pi=990$~MHz,
and $g_b^a/2\pi=980$~MHz. In this case, the energy transfer can be accomplished within about $300$~ns.
Figure~\ref{figure2}(b) simulates the energy transfer dynamics in the moderately-clustered geometry
with $J_{12}=J_{34}=1.62J_{23}$. Here, we adopt $g_{1}/2\pi=g_{4}/2\pi=150$~MHz,
$g_{2}/2\pi=g_{3}/2\pi=990$~MHz, and $g_b^a=930$~MHz. Compared to Fig.~\ref{figure2}(a), the energy
transfer from the first qubit to the last one can be completed within a shorter time, i.e.,
approximately $150$~ns. The moderately-clustered geometry supports a faster energy transfer because
the enhanced couplings within the cluster enlarge the intra-cluster energy gap and reduce the
inter-cluster energy gap. Both the strong coherent hopping within the cluster and the resonant energy
transfer between the two clusters accelerate the overall energy transfer. Moreover, when the ratio
$J_{12}/J_{23}$ increases to a larger value, e.g., $J_{12}=J_{34}=3.11J_{23}$ for $g_{1}/2\pi=g_{4}/2\pi=230$~MHz,
$g_{2}/2\pi=g_{3}/2\pi=920$~MHz, and $g_{b}^{a}=800$~MHz, the energy transfer becomes extremely slow
and it does not finish even at 300~ns, as shown in Fig.~\ref{figure2}(c). That's because the strong
intra-cluster couplings enlarge the intra-cluster energy gap excessively and thus increase the
inter-cluster energy gap. In order to find out the optimal parameters for the EET efficiency, we simulate
the EET dynamics of our circuit for a broad range of the parameters $g_1$, $g_2$, and $g_b^a$ with $g_{1}=g_{4}$
and $g_{2}=g_{3}$ and keeping other parameters unchanged. The parameters are changed from $10$~MHz
to $990$~MHz with $10$~MHz step. According to the numerical simulations, we find that the optimal energy
transfer occurs at $J_{12}=J_{34}=1.62 J_{23}$, as shown in Fig.~\ref{figure2}(b).

\begin{figure}[!h]%[tpb]  % figure 4
\begin{center}
\includegraphics[width=8.5 cm,angle=0]{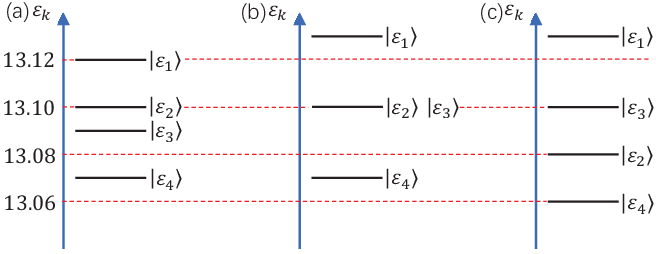}
\caption{ The energy-level diagrams corresponding to quantum dynamics in Fig.~\ref{figure2}:
(a) equally-spaced geometry with $J_{12}=J_{34}\approx J_{23}$ and $r\approx R/3$, (b) moderately-clustered geometry with $J_{12}=J_{34}=1.62J_{23}$ and $r<R/3$, (c) over-clustered geometry with $J_{12}=J_{34}=3.11J_{23}$ and $r\ll R/3$. $r$ and $R$ are respectively the intra-pair distance and the distance between the two ends in Fig.~\ref{figure1}(b).
}\label{figure3}
\end{center}
\end{figure}

In Fig.~\ref{figure2}, there are coherent oscillations in the short-time regime, which correspond to
the quantum coherence phenomena discovered in 2D spectroscopy experiments \cite{Cheng09,Pachon12}.
Because there is strong coupling between the two donor (acceptor) qubits, the energy gap between their
eigen-states will be opened up, i.e., $\vert\varepsilon_1\rangle$ and $\vert\varepsilon_2\rangle$, as
shown in Fig.~\ref{figure1}(b). And the energy coherently moves within the donor cluster. However, since
the couplings between the donor and acceptor clusters are relatively weak, the energy will incoherently
hop between the clusters. According to F\"{o}rster theory \cite{Valkunas13,May11}, the energy transfer
rate from the lower donors' eigen-state $\vert\varepsilon_2\rangle$ to the higher acceptors' eigen-state
$\vert\varepsilon_3\rangle$ is integral of donor's emission spectrum and acceptor's absorption spectrum.
The former is centered at the lower donors' eigen-state, and the latter is centered at the higher acceptors'
eigen-state. When the coupling within the donor (acceptor) pair is moderately enlarged,
cf. Fig.~\ref{figure3}(a)~vs~(b), the energy gap between $\vert\varepsilon_2\rangle$ and $\vert\varepsilon_3\rangle$
has been reduced, and thus results in an enhanced transfer rate. In this way, the clustered geometries
utilize energy matching to optimize energy transfer between two clusters. However, if the coupling within
the donor (acceptor) pair is over-enlarged, cf. Fig.~\ref{figure3}(c), this may suppress the transfer
rate as the energies of two states mismatch \cite{Mancal13}. Moreover, strong coupling between a charge
qubit and a TLR can be achieved in superconducting circuit \cite{Ashhab10,Fay08,Nataf10}.

\begin{figure}[!h]%[tpb]  % figure 3
\begin{center}
\includegraphics[width=8.5 cm,angle=0]{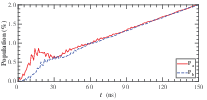}
\caption{ The populations trapped in the TLRs $P_{a}$ ($P_{b}$) vs time.}
\label{figure4}
\end{center}
\end{figure}

In Fig.~\ref{figure2}, we also show the quantum simulations by $H_\textrm{eff}$. The simulations by $H_\textrm{eff}$ coincide with those by $H_\textrm{I}$ in the aspects of oscillation frequency and energy transfer rate.
We notice that in Fig.~\ref{figure2}(b), the population of each qubit is about $24\%$, when the circuit is in the steady-state. To explore the reason why the summation of all the populations of the four qubits can not reach $100\%$, we plot the time evolution of populations in the TLRs $P_{a}=\langle1_{a}\vert\rho\vert1_{a}\rangle$
and $P_{b}=\langle1_{b}\vert\rho\vert1_{b}\rangle$ in Fig.~\ref{figure4}. It is shown that after some oscillations in the short-time regime, $P_{a}$ and $P_{b}$ increase linearly in time. And both of them reach about $2\%$ at 150~ns. Therefore, since only a small portion of the population, which has not been transferred to the target qubit, has been trapped in the two TLRs,  our simulation approach can effectively mimic the EET in the 4-chromophore linear photosynthetic complex, as shown in Fig.~\ref{figure1}(b).

\begin{figure}[!h]%[tpb]  % figure 4
\begin{center}
\includegraphics[width=8.5 cm,angle=0]{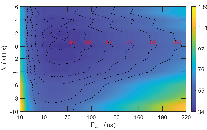}
\caption{ The transfer time $\tau$ vs the detuning $\Delta=\varepsilon_2-\varepsilon_3$
between the lower eigen-state $\vert\varepsilon_2\rangle$ of $\vert1\rangle$ and $\vert2\rangle$, and
the higher eigen-state $\vert\varepsilon_3\rangle$ of $\vert3\rangle$ and $\vert4\rangle$, and the
dephasing time $\Gamma_{\phi}^{-1}$.
}\label{figure5}
\end{center}
\end{figure}

Figures~\ref{figure2} and~\ref{figure3} clearly show the dependence of transfer time $\tau$ on the
detuning $\Delta=\varepsilon_2-\varepsilon_3$ between the lower eigen-state $\vert\varepsilon_2\rangle$
of $\vert1\rangle$ and $\vert2\rangle$, and the higher eigen-state $\vert\varepsilon_3\rangle$ of
$\vert3\rangle$ and $\vert4\rangle$. And the previous discoveries \cite{Mostame12,Gorman18}
suggested the design of bath can also optimize the energy transfer. Therefore, we investigate the
energy transfer for a broad range of $\Delta$ and the dephasing time $\Gamma_{\phi}^{-1}$,
as shown in Fig.~\ref{figure5}. By using a two-exponential-decay model in Ref.~\cite{Ai13},
\begin{equation}
P_4(t)=a_L(1-e^{-t/\tau})+a_S(1-e^{-t/\tau^{\prime}}),
\end{equation}
where $a_L\gg a_S$,
we numerically fit the population dynamics of qubit 4, and effectively obtain the transfer time $\tau$.
We may identify the optimal transfer time $\tau_\textrm{opt}=35.03$~ns at $\Delta=0$ and $\Gamma_{\phi}^{-1}=60$~ns.
First of all, we explore the relation between $\tau$ and $\Delta$, for a fixed $\Gamma_{\phi}^{-1}$.
When $\Delta$ is decreased from a large positive detuning, the transfer time experiences a decrease as
$\vert\varepsilon_2\rangle$ and $\vert\varepsilon_3\rangle$ approach the energy matching, i.e., $\Delta=0$.
However, if $\Delta$ further decreases, the transfer time increases since the two states become more and
more mismatched. On the other hand, we also explore the relation between $\tau$ and $\Gamma_{\phi}^{-1}$,
for a fixed $\Delta$, i.e., in a given geometry. By tuning $\Gamma_{\phi}^{-1}$, the system gradually reaches
optimal transfer time at $\Gamma_{\phi}\simeq16.7$~MHz, which is comparable to the effective coupling $J_{23}=8.87$~MHz
between $\vert2\rangle$ and $\vert3\rangle$. In the coherent-dynamics limit, i.e., $\Gamma_{\phi}^{-1}\rightarrow\infty$,
because the population will oscillate backwards and forwards, the energy can not be effectively transferred.
In the opposite case, when $\Gamma_{\phi}^{-1}\rightarrow0$, since the strong system-bath couplings frequently
probe the qubits' populations, the dynamic localization freezes the population dynamics \cite{Plenio08,Rebentrost09,Chin10}.
In other words, the quantum Zeno effect \cite{Kofman12,Ai13-2} prohibits the effective energy transfer.

\section{DISCUSSION AND SUMMARY}
\label{sec:summary}

In this paper, instead of direct couplings among two neighbouring qubits, we explore two
additional TLRs to induce the couplings among any two qubits. In linear geometries, there
are next-nearest-neighbor couplings besides the nearest-neighbor couplings. This architecture
also enables us to simulate photosynthetic complexes beyond the linear geometries in Ref.~\cite{Ai13},
since natural photosynthetic complexes possess more interesting geometries, such as ring-shape
LH1(LH2), FMO, and photosystem I(II). Furthermore, by using TLRs, we can simplify the quantum
circuit for non-linear geometries beyond the nearest-neighbor couplings.

In our simulations, we select the charge qubits with the dissipation time $T_{1}=200$~$\mu$s \cite{Kim11},
which is much longer than their pure-dephasing time $T_{2}=0.07$~$\mu$s \cite{Kim11}, because
in dephasing-assisted photosynthetic energy transport the spontaneous fluorescence can be ignored
\cite{Plenio08,Rebentrost09,Chin10}. However, since the energy transfer generally completes within
$0.3~$$\mu$s, e.g. in Figs.~\ref{figure2}(a,b), a much shorter dissipation time, e.g. $T_1=3~\mu$s,
is enough to obtain the same simulations in realistic experiments. This has been confirmed by our
numerical simulations which are not shown here. Furthermore, due to couplings to the TLRs, the qubits'
decay times may be shortened due to Purcell effect. In Ref.~\cite{Sete14}, the Purcell decay rate
is analytically estimated as
\begin{equation}
\Gamma=\frac{\kappa}{2}-\frac{\sqrt{2}}{2}\sqrt{-A+\sqrt{A^2+(\kappa\Delta)^2}},
\label{eq:PurcellRate}
\end{equation}
where
\begin{equation}
A^2=\Delta^2+4g^2-\kappa^2/4,
\end{equation}
$\kappa^{-1}$ is the decay time of the TLR, $\Delta$ is the detuning between the qubit and TLR,
$g$ is the coupling strength. Because the qubits dispersively couple with the TLRs, i.e.,
$g/\Delta\ll1$, $\Gamma\simeq\kappa g^2/\Delta^2$ to the lowest order of $g/\Delta$. For the present
parameters, the Purcell decay time is about 2 ms, which is much longer than $4~\mu$s. Therefore,
the Purcell effect will not significantly modify the qubits' decay times.

Parameters used here have been realized in experiments. The quality factor of a superconducting
TLR can reach $10^5$ \cite{Megrant12}. The frequency of the fundamental mode of the TLR can be
designed from $1$~GHz to $10$~GHz \cite{Blais07,LaHaye04,Gaidarzhy05}. The frequency of a
superconducting charge qubit can be effectively tuned from $5$~GHz to $15$~GHz, by varying the
flux that applied though the loop of the qubit \cite{Blais07}. In addition, the ultra-strong
coupling between a charge qubit and a TLR is achieved when $\min\{g_{j}\}\gg\sqrt{\kappa_\alpha\Gamma_{j}^{\gamma}}$ ($j=1,2,3,4$, $\alpha=a,b$) \cite{Haroche94}.
Although the coupling strength reaches the ultra-strong regime, i.e., $g_j/\omega_a\simeq1/3$,
the Jaynes-Cumming model is still valid as the large-detuning condition still holds \cite{Kockum19}.
As a result, the population dynamics will not be essentially modified.

In Ref.~\cite{Wang18}, the NMR experimental simulation was compared to the numerically-exact
simulation using the hierarchical equation of motion (HEOM) \cite{Ishizaki09,Tao20}. Here, in our configuration,
there are 2 TLRs in addition to 4 qubits. The presence of extra modes in the TLRs results in
the increasing complexity  of the quantum master equation, which can be solved by using QuTiP \cite{Johansson13,Johansson12},
as compared to the HEOM for 4 chlorophylls.

In natural photosynthetic complexes, the energy is transferred from the outer antenna to the
reaction center across tens of nanometers \cite{Cheng09,Ai13,Wang18}. There is a large energy
gap between the lowest eigen-state of the outer antenna and the reaction center, which can
prevent the back transfer of energy. As a result, the transfer rate from the lowest-energy state to
the reaction center is much smaller than the transfer rate within the outer antenna \cite{Wang18,Moix12}.
 In this work, it is thus reasonable to simulate the energy transfer without the reaction center.
However, a TLR can be utilized to effectively simulate the reaction center. Coupled to some
specific qubit, the TLR can act as a reaction center through Purcell-like coupling \cite{Potocnik17,Chin18}.
By tuning the TLR's decay time and the qubit-TLR detuning, we can effectively modified the
qubit's decay time through Purcell-like coupling, according to Eq.~(\ref{eq:PurcellRate}).

In summary, we have proposed a simulation scheme for demonstrating geometric effects on the
photosynthetic EET in four superconducting charge qubits plus two separated high-$Q$ TLRs.
The loss of population during the EET is trapped in the TLRs. In the future work, it might
be interesting to demonstrate the effect of fluorescence on the EET by varying the couplings
between the qubits and the TLRs.

\section*{ACKNOWLEDGMENTS}

This work was supported by National Natural Science Foundation of China
under Grant Nos. 11674033, 11474026, 11505007, and Beijing Natural
Science Foundation under Grant No. 1202017.
M.H. was supported by the National Natural Science Foundation of China
under Grant Nos. 11647042 and 11704281.
M.J.T. was supported by the China Postdoctoral Science Foundation
under Grant No. 2018M631438.

\section*{Appendix: Complete Expression of $H_{2}$}
\label{SG}

\setcounter{equation}{0}
\renewcommand{\theequation}{A\arabic{equation}}

\setcounter{table}{0}
\renewcommand{\thetable}{A\Roman{table}}

Here, we give the detailed expression of $H_{2}$ used in Sec.~\ref{sec:scheme}. We can omit the high-order terms of $g_{j_{s}}/\delta_{j_{s}}$ and simplify
\begin{eqnarray}\label{eq3}     %  equation 3
H_{2}&=&U^{\dagger}H_{1}U
\end{eqnarray}
as
\begin{eqnarray}\label{eq4}             %  equation 4
H_{2}&=&\omega_{a}a^{\dagger}a+\omega_{b}b^{\dagger}b+g_{a}^{b}\left(a^{\dagger}+a\right)\left(b^{\dagger}+b\right)         \nonumber\\
&&      +\sum_{j_1=1}^{2}\frac{g_{a}^{b}g_{j_1}^{2}}{2\delta_{j_1}^{2}}\left(a^{\dagger}+a\right)\left(b^{\dagger}+b\right)\sigma_{j_1}^{z}
        \nonumber\\
&&+\sum_{j_2=3}^{4}\frac{g_{a}^{b}g_{j_2}^{2}}{2\delta_{j_2}^{2}}\left(a^{\dagger}+a\right)\left(b^{\dagger}+b\right)\sigma_{j_2}^{z}           \nonumber\\
&&+\sum_{j_1=1}^{2}\left[\frac{\omega_{j_1}}{2}\sigma_{j_1}^{z}+\frac{g_{j_1}^{2}}{\delta_{j_1}}
            \left(aa^{\dagger}\sigma_{j_1}^{+}\sigma_{j_1}^{-}-a^{\dagger}a\sigma_{j_1}^{-}\sigma_{j_1}^{+}\right)  \right. \nonumber\\
&&\left.+\frac{g_{a}^{b}g_{j_1}}{\delta_{j_1}}\left(b^{\dagger}\sigma_{j_1}^{-}+b\sigma_{j_1}^{+}\right)\right]
            +\frac{g_{1}g_{2}}{2\delta_{1}\delta_{2}}\left(\delta_{1}+\delta_{2}\right)\left(\sigma_{1}^{-}\sigma_{2}^{+}+\sigma_{1}^{+}\sigma_{2}^{-}\right)          \nonumber\\
&&+\sum_{j_2=3}^{4}\left[\frac{\omega_{j_2}}{2}\sigma_{j_2}^{z}+\frac{g_{j_2}^{2}}{\delta_{j_2}}
            \left(bb^{\dagger}\sigma_{j_2}^{+}\sigma_{j_2}^{-}-b^{\dagger}b\sigma_{j_2}^{-}\sigma_{j_2}^{+}\right)  \right. \nonumber\\
&&\left.+\frac{g_{a}^{b}g_{j_2}}{\delta_{j_2}}\left(a^{\dagger}\sigma_{j_2}^{-}+a\sigma_{j_2}^{+}\right)\right]
            +\frac{g_{3}g_{4}}{2\delta_{3}\delta_{4}}\left(\delta_{3}+\delta_{4}\right)\left(\sigma_{3}^{-}\sigma_{4}^{+}+\sigma_{3}^{+}\sigma_{4}^{-}\right)          \nonumber\\
&&+\frac{g_{a}^{b}g_{1}g_{3}}{\delta_{1}\delta_{3}}\left(\sigma_{1}^{-}\sigma_{3}^{+}+\sigma_{1}^{+}\sigma_{3}^{-}\right)
            +\frac{g_{a}^{b}g_{1}g_{4}}{\delta_{1}\delta_{4}}\left(\sigma_{1}^{-}\sigma_{4}^{+}+\sigma_{1}^{+}\sigma_{4}^{-}\right)     \nonumber\\
&&+\frac{g_{a}^{b}g_{2}g_{3}}{\delta_{2}\delta_{3}}\left(\sigma_{2}^{-}\sigma_{3}^{+}+\sigma_{2}^{+}\sigma_{3}^{-}\right)
            +\frac{g_{a}^{b}g_{2}g_{4}}{\delta_{2}\delta_{4}}\left(\sigma_{2}^{-}\sigma_{4}^{+}+\sigma_{2}^{+}\sigma_{4}^{-}\right).
\end{eqnarray}

\end{document}